\documentclass[12pt]{iopart}
\usepackage{epsfig}
\usepackage{graphicx}
\usepackage{dcolumn}
\usepackage{bm}
\usepackage{iopams}

\begin{document}

\title[Bound dimers in bilayers of cold polar molecules]{Bound dimers in bilayers of cold polar molecules}

\author{A G Volosniev$^{1}$, N T Zinner$^{1,2}$, D V Fedorov$^{1}$, A S Jensen$^{1}$ and B Wunsch$^{3}$}

\address{                    
  $^{1}$Department of Physics and Astronomy - Aarhus University, Ny Munkegade, bygn. 1520, DK-8000 \AA rhus C, Denmark\\
  $^{2}$The Niels Bohr Institute, Blegdamsvej 17, DK-2100 Copenhagen \O, Denmark\\
	$^{3}$Department of Physics, Harvard University, 17 Oxford Street, Cambridge MA, 02138, USA
}

\date{\today}

\begin{abstract}
The exploration of cold polar molecules in different geometries is a
rapidly developing experimental and theoretical pursuit. Recently, the
implementation of optical lattices has enabled confinement in stacks
of planes, the number of which is also controllable. Here we consider
the bound state structure of two polar molecules confined in two
adjacent planes as function of the polarization angle of the dipole
moment of the molecules. We prove analytically and present numerical
evidence for the existence of bound states for arbitrary dipole
moments and polarization directions in this two-dimensional geometry.  The
spatial structure of the bound states is dominated by two-dimensional
$s$- and $p$-waves, where the latter exceeds 40 percent over a large
range of polarization angles for intermediate or strong dipole
strength.  Finally, we consider the influence of the dimer bound
states on the potential many-body ground-state of the system.
\end{abstract}

\maketitle

\section{Introduction}
A strong experimental drive in the field of polar atoms and molecules
has realized controllable samples in the rotational and vibrational
ground-state that are close to quantum degeneracy
\cite{ospelkaus2008,ni2008,deiglmayr2008,lang2008,ni2010,ospelkaus2010}.
These heteronuclear systems have a number of very interesting
properties due to the long-range and anisotropic dipole-dipole force
which can give rise to highly non-trivial many-body states in both the
weak- and strong-coupling regime \cite{baranov2008,lahaye2009}.  The
attractive head-to-tail configuration can, however, lead to strong 
chemical reactions \cite{ospelkaus2010} or many-body collapse
of the system \cite{lushnikov2002}, and confinement in optical
lattices has been suggested as a means of avoiding this problem
\cite{wang2006}. These confined one- or two-dimensional geometries
have led to a number of predictions of novel few- and many-body states
\cite{wang2006,wang2007,buchler2007,wang2008,bruun2008,sarma2009,cooper2009,
sun2010,yamaguchi2010,pikovski2010,zinner2010,zinner2011,levinsen2011},
and very recently the first experimental implementation of a
multilayered stack of pancakes containing fermionic polar molecules
was reported \cite{miranda2010}.

Here we consider the case of two adjacent layers. However, even in
this seemingly simple case there is a competition of intra- and
interlayer interactions which can vary between repulsion and
attraction as one changes the polarization angle of the dipole moments
with respect to the layers.  In the present paper we will be concerned
with few-body states with one particle in each layer in order to
describe the simplest complex in such a system in detail.  The case of
dipoles oriented perpendicular to the layers was considered from the
few-body bound state and scattering point of view in previous works
\cite{shih2009,ticknor2009,jeremy2010,klawunn2010}.  At the so-called
'magic' angle where the intralayer repulsion vanishes in a
one-dimensional trap the few-body bound state structure was also
discussed \cite{santos2010,deu2010,wunsch2011}. 
 
To our knowledge, the full
two-body bound-state problem as a function of the polarization angle
and the dipole moment has not been studied previously.
This problem is highly non-trivial due
to (i) the anisotropy and (ii) the vanishing integral over space of
the potential for arbitrary polarization angle.  The problem is a
specification of a more general problem of two particles in two
dimensions interacting via anisotropic potentials where the net volume
is negative or zero. The more general problem has been addressed only in a recent letter
\cite{vol11}, where the details of analytical and numerical methods are 
not elaborated.

In this paper, we present in section 2 analytic results for the
dipole-dipole potential. They are specialization of the derivation in
Appendix A for an arbitrary potential.  In section 3, we describe the
novel numerical method based on stochastic variation along with the
computed results for the dipole-dipole potential.  We present
energies, wave functions, and expectation values of relevant operators
as functions of the potential strength and the polarization angle.
One of our main results, analytical as well as numerical, is that the
bilayer system has a bound state for {\it any} polarization angle and
{\it any} value of the dipole moment.  We also calculate a
partial-wave decomposition that characterizes the geometric structure
of the wave function which indicates the likely symmetries of the
corresponding many-body problem. In section 4 we present a first
application of our results in a many-body context. We consider the
limit of strong coupling where the system forms bound bosonic dimers
that can potentially form a (quasi)-condensate. Finally in section 5
we brifly summarize and conclude.

\section{Analytic results}
The general setup we consider consists of two particles confined to two spatial
dimensions with a pair potential $V(\vec r)$ depending on the relative
coordinate $\vec r$. In polar coordinates, $\vec r = (x,y) = (r\cos
\varphi, r\sin \varphi)$, we have the Schr\"{o}dinger equation:
\begin{equation} \label{e30}
\left[-\frac{1}{s}\frac{\partial}{\partial
s}{s}\frac{\partial}{\partial
s}- \frac{1}{s^2}\frac{\partial^2}{\partial \varphi^2} +
 \lambda \bar V(s,\varphi)\right]\Psi=\alpha^2\Psi \;,
\end{equation}
where $\Psi$ is the wave function, $\mu$ the reduced mass, $\lambda$
is the dimensionless strength, $\lambda\bar V = 2\mu d^2 V/\hbar^2$ and
$\alpha^2=2\mu d^2E/\hbar^2$, $V$ is the potential and $E$ the energy, $d$
the unit of length, and $s=r/d$, is the reduced coordinate.

The investigation of possible bound states was briefly sketched
analytically in \cite{vol11} for an arbitrary potential in two spatial
dimensions (2D).  The result is that any cylindrical or non-cylindrical
potential has a least one bound state provided the volume of the
potential is negative or zero.  We give more details of a similar
derivation in Appendix A.  The difficulties are centered around
exceedingly weak potentials where no ordinary perturbation treatment is
applicable because no unperturbed solution is available. The binding
energy approaches zero as the potential vanishes and the continuum is
approached.  Thus the limiting energy is zero and the corresponding
wave function is uniformly distributed over all space.

\begin{figure}[thb]
\centerline{\input{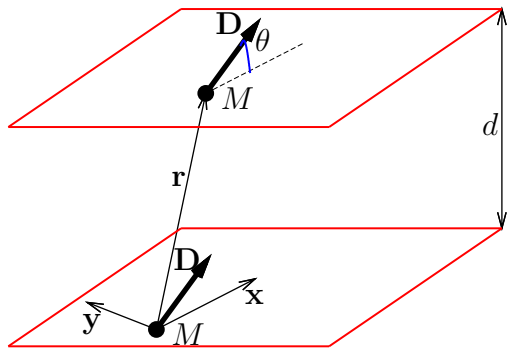}}\vspace{1cm}
\caption{Illustration of the setup consisting of two dipolar particles of mass $M$ moving in 
parallel planes separated by a distance $d$. Their dipole moments, $\bm D$, are assumed to 
be aligned by an external field at angle $\theta$ with respect to the planes.} \label{fig-setup}
\end{figure}

We specialize to the dipole-dipole potential arising for the
system of two polarized molecules of mass $M$ confined
to two parallel planes separated by a distance $d$ as shown in 
figure \ref{fig-setup}. The corresponding
dipole-dipole potential, $V$, projected to this two-dimensional
geometry is
\begin{equation} \label{e20}
 V(r,\varphi)=D^2 \frac{r^2+d^2-3(r\cos\varphi\cos\theta+
 d\sin\theta)^2}{(r^2+d^2)^{5/2}}\;,
\end{equation}
where $D$ is the dipole moment \footnote{In SI units we have
  $D^2=d^2/4\pi\epsilon_0$ when $d$ is dipole moment of the molecules.
  However, in this paper we use $d$ to denote the interlayer
  distance.}, and $\theta$ denotes the polarization angle measured
from the layer plane to the $z$-axis which intersects the two layers
at right angles.  

The potential in equation (\ref{e20}) is found in the ideal limit of
zero-width layers.  The dipole polarization is measured such that for
$\theta=\pi/2$, the dipoles are oriented perpendicular to the layers
as in \cite{shih2009,jeremy2010,klawunn2010}.  One can take
corrections to the zero-width layer limit into account by integrating
out a gaussian in the transverse direction. However, the corrections
are second-order in the width, $w$, and we neglect them as we here are
interested in the $w\ll d$ limit.

We have to solve the 2D Schr\"odinger equation in equation (\ref{e30}) with
the potential in equation (\ref{e20}). The reduced mass is $\mu=M/2$
and the dimensionless dipolar strength is given by $U= MD^2/(\hbar^2d)$,
which is a measure of the ratio of potential to kinetic energy. 
We will also
consider the case where $U<0$ which is also physically realizable as
explained below. In the notation above we have
$\lambda=U$ and we will use these notations interchangeably in 
order to emphasize the generality of the analytic approach presented
in Appendix A.

This potential is invariant under reflection in the $x$-axis and has
the peculiar property that $\int d\,xd\,y V(x,y)=0$ for any $\theta$.
In particular, it does not fulfil the Landau criterion for bound
states in two dimensions \cite{landau1977}, which states that a bound state
always exists for $\int d\,xd\,y V(x,y)<0$. 
An early existance proof was given in
\cite{simon1976} using a method that is not well-suited for
expansions in the strengh of the potential. A discussion of such an 
expansion appeared in \cite{klawunn2010} but only for case
where $\theta=\pi/2$ and cylindrical symmetry holds.
Here we are interested in the appearance and
properties of bound states for arbitrary $\theta$.  A partial-wave
decomposition of the potential in the basis $\{1,\cos\varphi,\cos
(2\varphi)\}$ (which are the only non-zero terms) leads to
\begin{eqnarray} \label{e39a}
&\lambda\bar V(s,\varphi) = V_0(s) + V_1(s)\cos\varphi 
+  V_2(s) \cos(2\varphi)&\\ \label{e39b}
&V_0(s)=U\frac{[3\sin^2\theta-1][s^2/2-1]}{(s^2+1)^{5/2}},&\\ \label{e39c}
&V_1(s)=-3U\frac{s \sin (2\theta)}{(s^2+1)^{5/2}},&\\ \label{e39d}
&V_2(s)=-\frac{3}{2}U\frac{s^2\cos^2\theta}{(s^2+1)^{5/2}},&
\end{eqnarray}
which we will refer to as monopole, dipole, and quadrupole terms,
respectively.  The monopole potential $V_0$, has in itself zero net
volume and it vanishes identically for $\theta=\theta_c$ where
$\sin^2\theta_c=1/3$.  The dipole term only vanishes for $\theta=0$ and
$\pi/2$, whereas the quadrupole term is finite except at
$\theta=\pi/2$. Thus for $\theta>\theta_c$ and $U>0$, the monopole
term has an inner attractive pocket and a repulsive barrier outside
$s=\sqrt{2}$, and vice versa for $\theta<\theta_c$. For $U<0$ the
story is reversed. We expect the monopole term to be most
important for the system properties, at least when it
is non-vanishing away from $\theta=\theta_c$.  However, the monopole
term is, except for the factor of $(3\sin^2\theta-1)/2$, identical to the
full potential at $\pi/2$, i.e. we know from previous work that it
always supports bound states \cite{jeremy2010,klawunn2010,simon1976}.
We also know that the configuration with an attractive inner pocket
and a repulsive outer barrier leads to considerably stronger binding
than in the reversed case \cite{jeremy2010}. We will see this
explicitly in the energies presented below.

It is very important to notice that the angle $\theta_{c}$ is different
from the magic angle, $\theta_{c}^{*}$, 
where the potential of two dipoles moving in one
dimension vanishes (determined by $\cos^2\theta_{c}^{*}=1/3$)
\cite{lahaye2009}. This demonstrates an important difference between
one- and two-dimensional dipolar systems. We will address this fact in
more detail when we discuss many-body physics below.

First we specialize the analytical
results in Appendix A derived for general interactions
to the dipolar potential.  A partial
wave expansion is employed in analogy to that of the decomposition in
equation (\ref{e39a}).  For the dipole potential in equation (\ref{e20}) the
resulting energy expression from equation (\ref{a9}) then becomes
\begin{equation} \label{e70}
  E  = - \frac{4\hbar^2}{M d^2} \exp\left(-2\gamma -
 \frac{2(1+U B_1+U^2B_2)}{U^2(A_0 + U A_1+U^2A_2)}\right) \; ,
\end{equation}
where $\gamma$ is Euler's constant and 
the coefficients, $A_0,A_1$ and $B_1,B_2$, are defined by
\begin{eqnarray} \label{e80}
A_0 &=& \frac{1}{4}M_c^2 +\frac{1}{8}
\sin^2{2\theta}+\frac{1}{32}\cos^4{\theta} \;, \\ 
A_1 &=& + 0.0053  
\sin^2{2\theta}\cos^2{\theta} - 0.0033 \sin{2\theta}\cos^4{\theta}
 \nonumber \\  &-& 0.0019 \cos^6\theta
- M_c\big(0.0349 \sin^2(2\theta)   \\ \nonumber
&+& 0.0054 \cos^4(\theta) 
+ 0.0156 M_c \cos^2\theta + 0.0343 M_c^2\big) \;, \\ 
B_1 &=& -  1.204 M_c - \frac{1}{16} \cos^2\theta \;, \\ 
B_2 &=&  0.8382 M_c ( M_c + 0.0667 \cos^2\theta) \\ \nonumber
  &-&  0.0037 \sin^2(2\theta) + 0.0894 \cos^4\theta \;,\\ 
M_c &=& \frac{3}{2}\sin^2(\theta)-\frac{1}{2} \;. \label{e90}
 \end{eqnarray}
The expression for $A_2$ is much more elaborate consisting of more
than a hundred terms each given as double, triple, and quadrupole integrals over well
defined functions. We refrain from showing them all here. Note that since
the spatial integral of the potential vanished, the term $A_2$ has to be considered
in an expansion to second order in $U$ (see also (\ref{a9})).

The only non-zero matrix elements for the dipole potential in
equation (\ref{e20}) are $V_{m,m}, V_{m,m\pm 1}, V_{m,m\pm 2}$.  This
implies that the $m=0$, 1, and 2  partial waves are sufficient to get all
contributions to the orders given on the right hand side of
equation (\ref{e70}). Higher partial waves beyond $m=2$ only contribute to the
wave function for the dipole potential through orders of $U$ that are higher than
those in equation \ref{e70}.

The scattering length, $a$, is usually considered to be a measure of
the most crucial model independent property of any potential.  In the
present case it is a function of strength and polarization angle, and
to any order related to the energy as in \cite{nie01}, i.e.  through
the general equation
\begin{equation} \label{e126} 
 E =  - 2 \frac{\hbar^2}{ \mu a^2} \exp(- 2 \gamma) \;.
\end{equation}
Any accuracy of $E$ is then
directly transferable to $a$ through equation (\ref{e126}).  The energy can
be calculated to any order in powers of $U$, as second order in
equation (\ref{e70}). Then the scattering length becomes
\begin{equation} \label{e136}
  \frac{a}{d} = 
 \exp\left(\frac{(1+U B_1+U^2B_2)}{U^2(A_0 + U A_1+U^2A_2)}\right) \; ,
\end{equation}
where the dependence on strength and polarization angle now is
explicit.

The energy, and scattering length, very close to threshold is
exponential in $U^{-2}$, as seen in eqs.(\ref{e70}) and (\ref{e136}),
and determined by the polarization angle through $A_0$.  The first
order terms, $(A_1,B_1)$, in $U$ exhibits the difference in approach
to threshold for the different signs of the strength, $U$.  The second
order terms, $(A_2,B_2)$, are necessary to get the correct
$U$-independent pre-exponential factor in the energy. Here $B_2$ is
both much simpler and more significant than $A_2$ which consists of
sums of a large number of contributions expressed as definite
integrals.

The expressions simplify substantially for $\theta =\pi/2$.  In
\cite{baranov2010} the energy is calculated for $\theta=\pi/2$ to the
order given in equation (\ref{e70}), including the $A_2$ term, in agreement
with our result. The computation of the energy in \cite{klawunn2010} for
$\theta=\pi/2$ deviates from our results in equation (\ref{e70}) in
the first order correction.

\section{Numerical procedure}
The potential is in general anisotropic and the wave equation is not
easy to solve by discretization or integration. We therefore turn to
the stochastic variational approach using gaussian wave functions
which has been successfully applied to other interactions
\cite{suzuki1998}.  However, in the limit of weak binding the wave
functions become very small and spatially extended without structure
at large distances. The special method to achieve convergence with a
fair amount of gaussians is described in this section, and the results for
energies and wave functions presented in the next section. 

\subsection{Method for weakly bound systems}
For numerical calculations we employ the correlated Gaussian method
which has been sucessfully used in a range of few-body problems in atomic
physics~\cite{martin7,martin8,marcelo10}.  The wave function $\Psi(x,y)$
is found through the variational principle as a linear combination of
basis-functions $G_i(x,y)$,
\begin{equation}
\Psi(x,y) = \sum_{i = 1}^{N_\mathrm{basis}} c_i G_i(x,y),
\end{equation}
where $N_\mathrm{basis}$ is the size of the basis, $c_i$ are the linear
variational parameters, and the basis-functions are chosen in the form of
shifted correlated Gaussians,
\begin{equation}
G_i(x,y)=e^{-(q-s_i)^TA_i(q-s_i)}\;,
\end{equation}
where the superscript $T$ denotes transposition, $q\equiv(x, y)^T$
is the column-vector of the coordinates, and where the elements of the
symmetric positive correlation matrices $A_i$ and the shift
vectors $s_i$ are the non-linear variational parameters. The explicit
shifts employed here enhance greatly the flexibility of the correlated
Gaussians specifically for the non-rotationally symmetric system at hand.

According to the variational principle the wave function is found by
minimizing the expectation value of the
Hamiltonian,
\begin{equation}
E[\Psi] \equiv
\frac{\langle\Psi|H|\Psi\rangle}{\langle\Psi | \Psi\rangle}.
\end{equation}
The linear parameters $c_i$ are determined by solving the generalized
eigenvalue problem
\begin{equation}\label{eq-generalized}
{\mathcal H}c=E[\Psi]{\mathcal N}c \;,
\end{equation}
where ${\mathcal H}$ and ${\mathcal N}$ are
$N_\mathrm{basis}\times N_\mathrm{basis}$ matrices,
\begin{equation}\label{eq-H-N}
{\mathcal H}_{ij}\equiv\langle G_i|H|G_j\rangle
\;\textrm{and} \;
{\mathcal N}_{ij}\equiv\langle G_i|G_j\rangle
\;,
\end{equation}
where $c$ is the column-vector of the linear parameters $c_i$. The
overlaps and the matrix elements of the kinetic energy operator in
equation (\ref{eq-H-N}) are calculated using the analytical expressions
in Appendix~B. The matrix elements of the interaction potential
are calculated numerically using adaptive Gauss-Kronrod
quadratures~\cite{GSL}.

The generalized eigenvalue problem equation (\ref{eq-generalized}) is solved
by first performing Cholesky decomposition of the (symmetric
and positive definite) matrix ${\mathcal N}=LL^T$,
and then solving the ordinary symmetric eigenvalue problem,
\begin{equation}
\tilde{\mathcal H}\tilde c=E[\Psi]\tilde c \;,
\end{equation}
where $\tilde{\mathcal H}=L^{-1}{\mathcal H}{L^{-1}}^T$ and $\tilde
c=L^Tc$, using standard linear algebra methods~\cite{GSL}.
The non-linear parameters -- the elements of the correlation matrices
$A_i$ and the shift vectors $s_i$ -- are determined through
stochastic sampling. Basically, the minimum of the energy is found by
sampling a large number of random sets of $N_\mathrm{basis}$ Gaussians
with randomly generated correlation matrices and shift vectors.  The
elements of the correlation matrices are generated as $\pm1/b^2$ and the
elements of the shift vectors as $\pm b$, where $b$ is a stochastic
variable with the dimension of length sampled from an exponential
distribution with certain length parameter $l$.
To ensure the positive definiteness of the correlation matrix the
following procedure is used: First a diagonal matrix with positive
elements $1/b^2$ is generated, and then a random orthogonal transformation
is performed. For ordinary quantum systems the length parameter, $l$, of the exponential
distribution should be chosen close to the typical range, $d$, of
the interaction potential between particles. For the dipolar 
potential considered in this paper, it is in fact the interlayer distance
$d$ which explains the abuse of notation. If the binding energy
$B$ of the system is of natural size, that is
$B\sim\frac{\hbar^2}{2\mu d^2}$, such sampling proves very effective.
However, for weakly bound systems, where $B\ll\frac{\hbar^2}{2\mu
d^2}$, two different length scales are of importance:
One is the interaction range $d$ and the other is the
typical size $|\alpha|^{-1}$ of the tail of the wave function (where
$\alpha=i\sqrt{\frac{2\mu B}{\hbar^2}}$). We therefore subdivide the basis
into $N_\mathrm{short}$, short-range, and $N_\mathrm{long}$, long-range,
Gaussians, $N_\mathrm{basis}=N_\mathrm{short}+N_\mathrm{long}$, where the
short-range Gaussians are sampled from the exponential distribution with
range $d$ and the long-range Gaussians are sampled from the exponential
distribution with range $|\alpha|^{-1}$. The latter is calculated from the
current estimate of the binding energy.

\begin{figure}[thb]
\centerline{\input{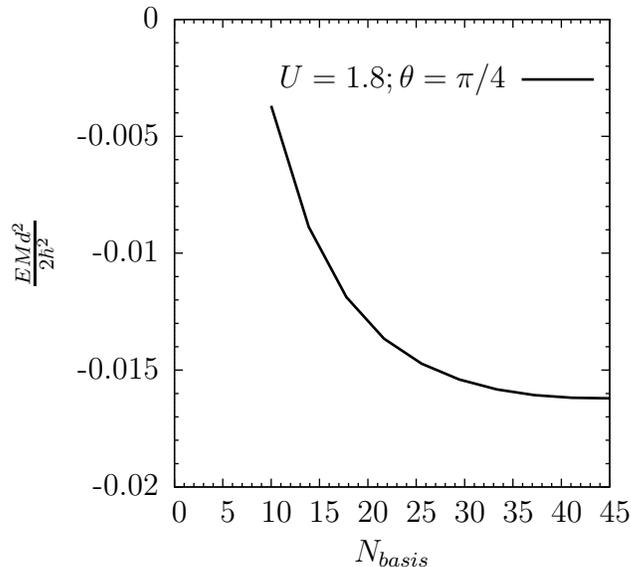}}\vspace{1cm}
\caption{Illustration of the convergence of the energy with respect to
the basis size: the energy $E$ of the ground state as function of the
size $N_\mathrm{basis}$ of the basis.} \label{fig-convergence-energy}
\end{figure}

\begin{figure}[thb]
\centerline{\input{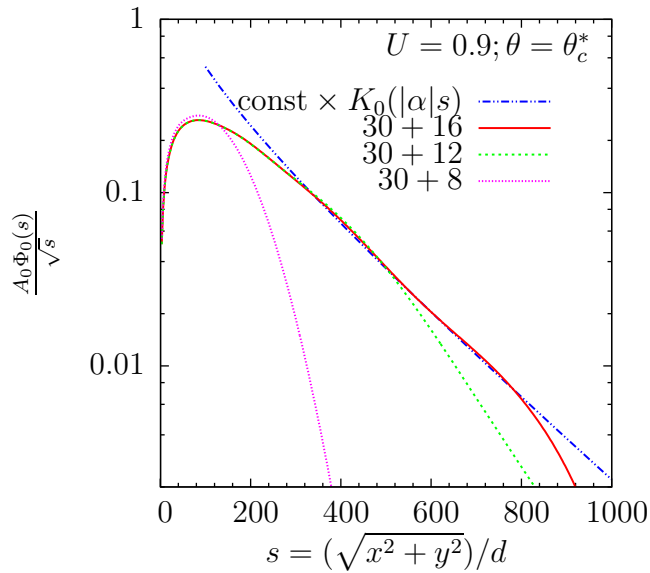}}\vspace{1cm}
\caption{Illustration of the convergence of the radial wave function
with respect to the basis size: the radial
function $\Phi_0(r)$ from equation (\ref{a4}) as function
of $s=\sqrt{x^2+y^2}/d$ calculated with bases of different sizes:
$N_\mathrm{short}=30$ and $N_\mathrm{long}=8,12,16$. For comparison
the analytic form of the tail, $K_0(|\alpha|r)$, is also shown.}
\label{fig-convergence-tail}
\end{figure}

Since the long-range Gaussians are introduced specifically to better
describe the asymptotics of the wave function, they can be chosen in a
much simpler form,
\begin{equation}\label{eq-long-range-gaussian}
G=e^{-r^2/b^2}\;,\; r\equiv\sqrt{x^2+y^2}\;.
\end{equation}
The typical convergence plot of the binding energy as function of the
basis size is presented in figure \ref{fig-convergence-energy}. It
shows that the energy is converged to within four significant digits
with the basis size of about 45 Gaussians.
The convergence of the tail of the wave function is illustrated
in figure \ref{fig-convergence-tail} where it is compared to the
analytic asymptotic form. Clearly, addition of the long-range Gaussians
significantly improves the quality of the wave function at very large
distances. In this example about 46 Gaussians, 30 short-ranged plus 16
long-ranged, can accurately describe the the wave function up to the
radius of about 800$d$.

\section{Results}
The general derivation in Appendix A demonstrates that there always is at
least one bound state in 2D for any anisotropic potential with zero
net volume as obtained in a different manner in \cite{simon1976}.  
When $U$ becomes small we expect universal behavior of
energies and radii \cite{vol11,nie01,jen04}.  Using the stochastic
variational approach in the small $U$ limit, our results for
$\theta=\pi/2$ approach the universal behavior of the energy which to
leading order scales like $\textrm{ln}(-E)\propto -1/U^2$ as
discussed previously \cite{jeremy2010,klawunn2010,vol11,simon1976}.
For other values of $\theta$ we expect the same scaling for very small
$U$, however, the range of $U$ around zero where this applies has a
strong dependence on $\theta$ as seen by the energies presented
below.  We also expect differences for general $\theta$ between
positive and negative $U$ in the limit $U\rightarrow0$ as for
$\theta=\pi/2$ \cite{jeremy2010}.  The question
of how the binding energy approaches universality is investigated in
more details in \cite{vol11}.

\begin{figure}
\centerline{\epsfig{file=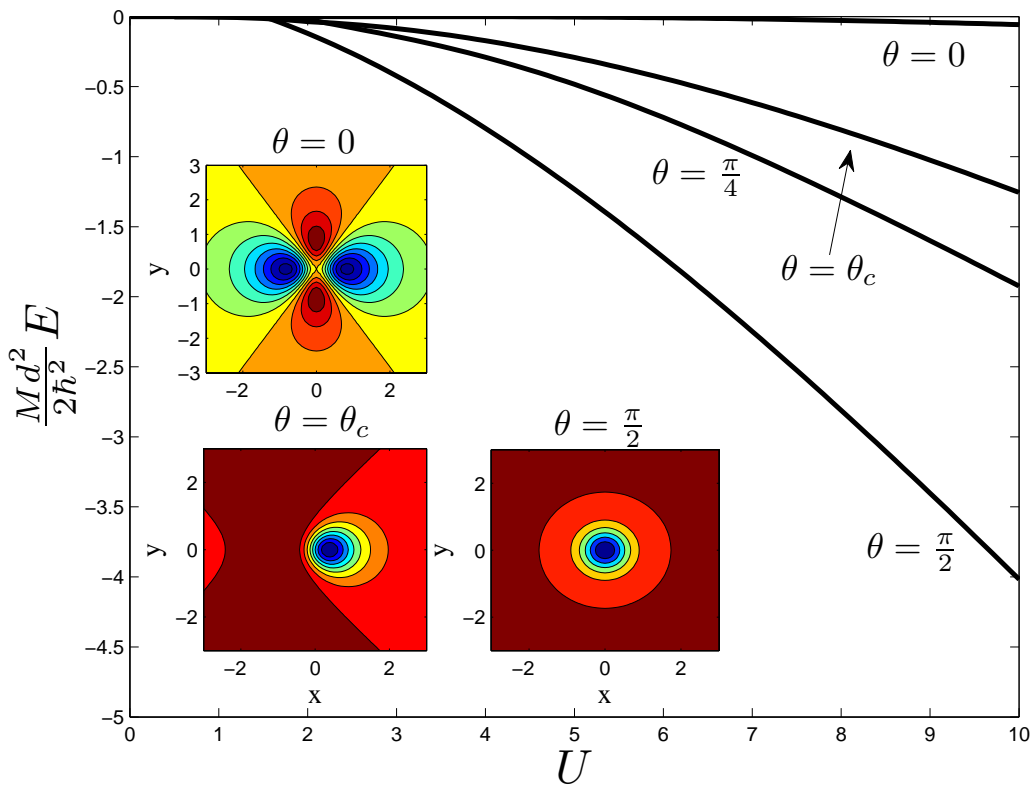,scale=0.75}}
\caption{Bound state energies as a function of $U$ for different
  angles. Insets show contour plots of the potentials with valleys in
  bright (blue) and hills in dark (red) colors. }
\label{fig-bound}
\end{figure}

The energies have been calculated using the correlated gaussian
approach. In figure \ref{fig-bound} we exhibit the results as a function
of $U>0$ for a selection of polarization angles.  At small $U$ the
energy decreases very fast with decreasing $U$ as noted already in
\cite{simon1976}, whereas at larger $U$ we find a linear dependence on
$U$ as argued in \cite{jeremy2010} for $\theta=\pi/2$.  The binding
energies decrease dramatically as $\theta$ approaches zero.  We stress
that we numerically find a bound state for any value of $U$ also in
the particularly unfavorable case of $\theta=0$ as well as for
$\theta_c$ ($\sin^2\theta_c =1/3$) where the decisive monopole
potential vanishes identically.  Notice that there are more bound
states for larger $U$ but we restricted our discussion to the single
bound state regime.

\begin{figure}
\centerline{\epsfig{file=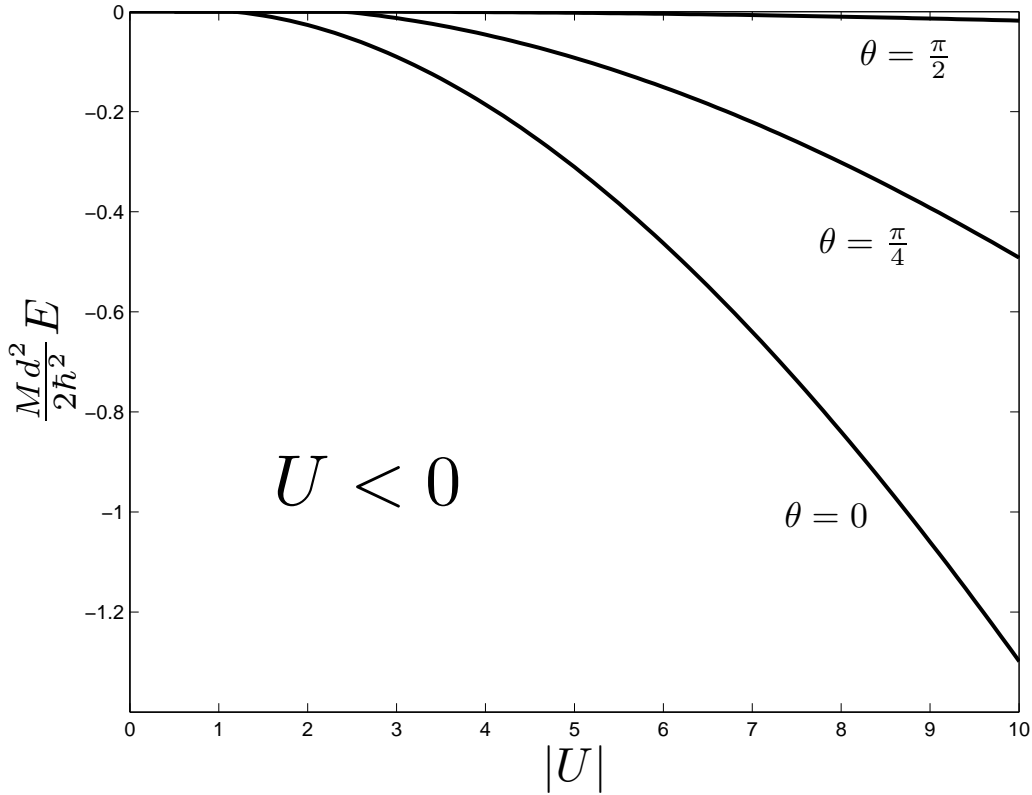,scale=0.75}}
\caption{Same as figure \ref{fig-bound} but for $U<0$. 
Note that the vertical scale is different from that of figure \ref{fig-bound}.}
\label{fig-neg}
\end{figure}

The $U<0$ case is also of great interest as that potential can be
generated by using microwave-dressed molecules. In
\cite{sarma2009,cooper2009,levinsen2011} an AC light field directed perpendicular
to the layers was used to create the $\theta=\pi/2$ potential with
$U<0$. A straightforward calculation shows that if the laser hits the
layers at an angle, $\theta$, the potential is the same as for a
homogeneous electric field at angle $\theta$ but with negative $U$.
For $U<0$ we again find numerically that for all values of the strength the
two-body system has bound states. The results for the binding energy
at different angles are shown in figure \ref{fig-neg} as function of
$|U|$. The first thing one notices is that the overall magnitude of
the bound state energy is smaller than that for $U>0$.  At
$\theta=\pi/2$ this can be understood as the potential has a repulsive
core at $\theta=0$, forcing the state to reside in the shallow
attractive pocket at intermediate distance.  In turn, this gives a
much smaller binding energy.  This qualitative behavior of the
potential persists until $\theta$ decreases below $\theta_c$ where the
monopole changes sign. Then the potential changes overall character to
become more attractive with inner attractive pocket and outer
repulsive tail. The $U<0$ results thus show maximum binding at
$\theta=0$ which is, however, still about a factor of three smaller
than the $U>0$ case at its most favorable angle of $\theta=\pi/2$.

\begin{figure}
\centerline{\input{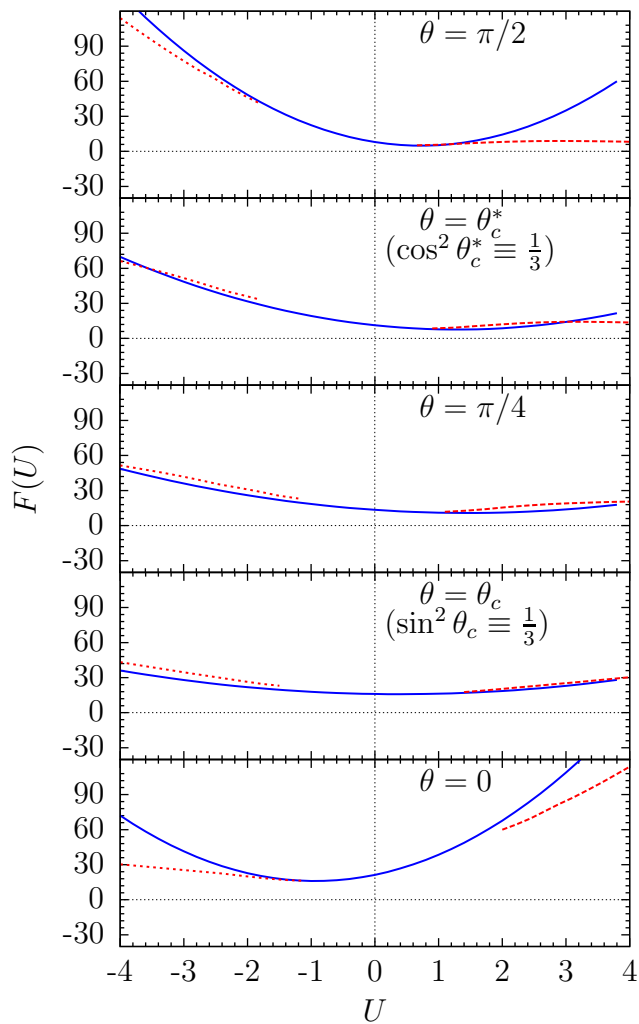}}\vspace{1cm}
\caption{The function $F(U) = -U^2\ln(|E|Md^2/(2\hbar^2))$ for
  different polarization angles $\theta$.  The dashed red curves are
  calculated numerically and the solid blue curves are from
  equation (\ref{e70}).  The relatively small contribution from the
  $A_2$-term is neglected in this comparison. }
\label{fig-exp}
\end{figure}

Numerical and analytical results are compared in figure \ref{fig-exp}.
To get the most accurate comparison we show the exponent in
equation (\ref{e70}) multiplied by the square of the strength, $U^2$. The
approach in the figure of analytical and numerical results is
reassuring in the limit of relatively weak potentials.  Since we
neglected the $A_2$-contribution this approach indicates its minor
significance.

\subsection{Wave functions}
\begin{figure}
\centerline{\epsfig{file=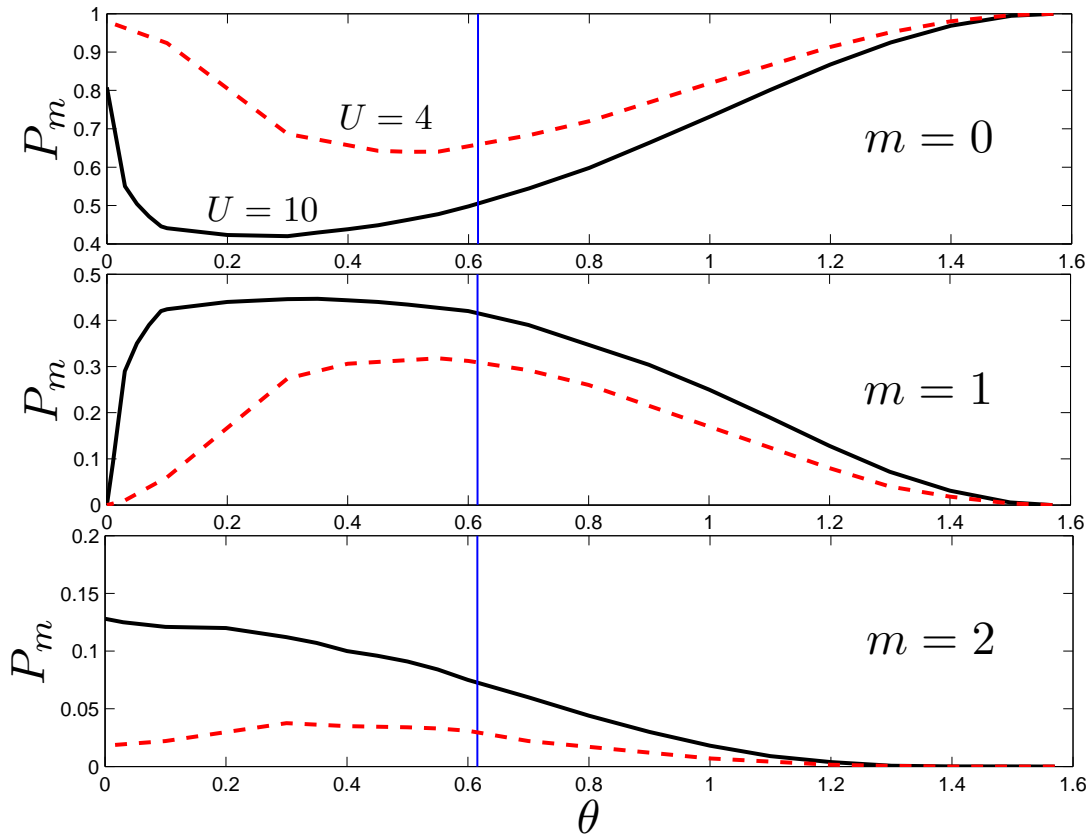,scale=0.75}}
\caption{Partial waves probabilities, $P_m=\pi(1+\delta_{m0})\int_{0}^{\infty}d\rho \rho|\Psi_m(\rho)|^2$ with
$\Psi(\rho,\phi)=\sum_m \Psi_m(\rho)\cos(m\phi)$, for the bound state wave 
function at $U=10$ (solid black) and $U=4$ (dashed red) as function 
of polarization angle $\theta$ for $m=0$, 1, and 2. The vertical line 
indicates $\theta=\theta_c$. Note the different vertical scales.}
\label{fig-part}
\end{figure}

The structure of the bound state wave functions can be seen from the
partial wave decomposition. The results are shown in
figure \ref{fig-part} for a strong coupling of $U=10$ and a weaker one
of $U=4$.  The probabilities are normalized so that they sum to
one. We note that the contribution of $m>2$ is only a few percent with
a maximum at $\theta=0$ of 5\% in $m>2$ terms. As expected we find
that $m=0$ becomes dominant for $\theta\rightarrow\pi/2$ as we
approach cylindrical symmetry. Interestingly, close to $\theta=0$ we
also find a very large $m=0$ component, no $m=1$ content, and a
significant $m=2$ contribution. The remaining content of the
wave function is found in the higher $m$ contributions. The fact that
$m=1$ has no weight for $\theta=0$ can be understood from the symmetry
of the potential. For $x\rightarrow -x$ the $m=1$ term changes sign,
whereas the potential is invariant.  Interestingly, as we go away from
$\theta=0$, the $m=1$ component raises rapidly and stays on the order
of 40\% until we reach $\theta=\theta_c$ at which it starts to decline
as for $m=2$, in line with the restoration of cylindrical symmetry at
$\theta=\pi/2$. For $U>10$ the $m=0$ component can be even more
suppressed in comparison to $m>0$ for intermediate $\theta$, whereas
for positive $U<4$ the $m=0$ component will eventually dominate as one
approaches the universal limit discussed above.

We have found similar results for the $U<0$ when taking into account
that the angle $\theta$ for $U>0$ correspond to angle $\pi/2-\theta$
for $U<0$ and vice versa.  This is in fact an exact symmetry of the
dipole part of the potential and an approximate one for the monopole
term since $\theta_c$ is close to $\pi/4$. For $U=-10$ we find that
there is a window $\theta_c<\theta<1.1$ in which the $m=1$ term is
around 40\%.  Interestingly, we find that the partial-wave content for
$U<0$ is almost exclusively $m=0$ and $m=1$. This is perhaps
surprising as the potential in the $m=2$ channel is non-vanishing
except at $\theta=\pi/2$.

\begin{figure}
\centerline{\input{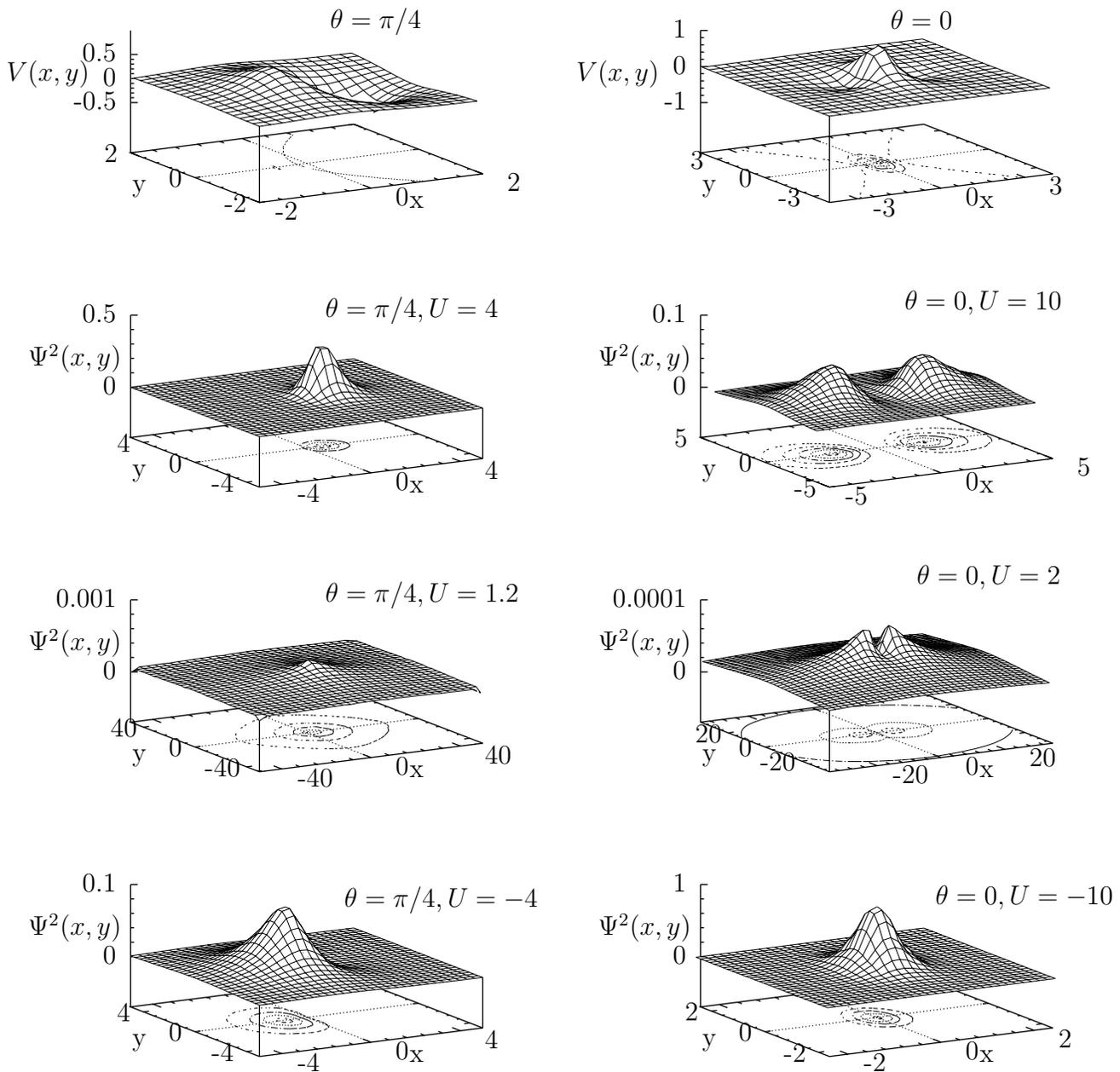}}\vspace{1cm}
\caption{Contour diagram for different polarization angles of the
  dipole-dipole potential divided by $U$ (upper part) followed by the 
  probability distributions below. All lengths are in units of $d$.}
\label{fig-potprob}
\end{figure}

The potential has completely different forms for different
polarization angles as illustrated on the contour plots in the inset 
of figure \ref{fig-bound}.  For $\theta =\pi/2$, 
the potential is cylindrical while asymmetry appears for
decreasing $\theta$-values, and eventually two minima emerge when
$\theta$ approaches zero.  For small $\theta$, the potential looks
like a harmonic oscillator along the $y$-axis for small $x$ and $y$.
The depth of this harmonic well around zero is about twice as large
for $\theta=0$ and $U<0$ in comparison to the depth of the two wells in
the $x$-direction for $\theta=0$ and $U>0$ that is shown in 
the inset of figure \ref{fig-bound} and figure \ref{fig-potprob}. A simple
gaussian wave function should therefore be a fair approximation to the
full problem and in turn the lowest partial-waves should dominate.

The wave functions for strongly bound states mimic the contours of the
potential.  As the strength, as well as binding energy, decreases the
wave function is spreading out to larger distances and approaching
cylindrical symmetry, as illustrated in figure \ref{fig-potprob}. The
probability decreases in all points of space and approaches zero
uniformly outside the potential.  However, inside the potential the
shape of the potential is maintained even for vanishingly small
strengths where the probability also approaches zero. This behavior is
necessary to provide binding which arises from the attractive
potential at small distances.  In turn, the modified Bessel 
function, $K_0(|\alpha|s)$, is
approached for vanishing strength which corresponds to a wavefunction
that is roughly constant in
space until the distance, $s$, is comparable to $1/|\alpha|$.

\section{The Many-Body Bilayer System}

The bilayer system has an interesting many-body structure with
combination of attractive interactions that can induce pairing and
repulsive interaction that tend to suppress such effects. This was
discussed recently for the $\theta=\pi/2$ case in
\cite{pikovski2010,zinner2010}. Here we consider the strongly-coupled
limit (large $U$) where the bound two-body dimers are expected to be
the relevant degrees of freedom. As the dimers are effectively bosons,
they are capable of forming a (quasi)-condensate under the right
conditions \cite{zinner2010}. However, as is well-known from BCS-BEC
crossover studies \cite{giorgini2008}, this is only expected to occur
when the density is low. Unfortunately, the
Berezinskii-Kosterlitz-Thouless (BKT) transition \cite{bkt1,bkt2} that
governs this two-dimensional system has a critical temperature that is
proportional to density \cite{pikovski2010,zinner2010}. Therefore, a
compromise where the dimer condensate occurs at not too low densities
would be optimal to allow experimental access to this unusual
many-body state.

A further interesting complication is the fact that each of the layers can 
accomodate various coherent many-body states when we consider them independently.
Proposals for the ground state include $p$-wave superfluids \cite{bruun2008,cooper2009} and
density waves \cite{sun2010,yamaguchi2010}. 
Of course, when more than one layer 
is present the long-range inter- and intralayer interactions are competing, and at this
point it is not clear what state is favoured for arbitrary directions of 
the polarization. It is known that the density-wave instability will be 
enhanced and occurs at smaller coupling strength for bi- and multilayer systems
\cite{zinner2011}. We note that in the $U<0$ bilayer case, a 
particle-hole coherence reminiscent of ferromagnetism has even been suggested \cite{sarma2009}.
Here we will be concerned mainly with the BCS-BEC crossover scenario in the 
strongly-coupled limit, but we will also 
estimate the appearance of a density-wave state and comment on possible superfluids.

As the criterion for the onset of condensation of dimers, we consider the point at which the chemical potential
becomes negative \cite{zinner2010}, i.e.
\begin{equation}
\mu(U,\theta)=\frac{1}{2}n V_{\textrm{eff}}(U,\theta)+E_F-\frac{1}{2}E_B(U,\theta),
\end{equation}
where $E_B$ is the dimer binding energy and $V_\textrm{eff}$ is the
long-wavelength (zero momentum) effective momentum-space interaction
between two dimers. Here we include both the binding energy and the
dimer-dimer interaction, and we also include a term for the Fermi
energy, $E_F$, that the constituents of the dimer inherit from their
layer.  The density of dimers (equal to the single-layer density when
the layers have an equal number of molecules) is denoted by $n$. To
obtain the effective interaction, one must in principle integrate out
the wave function of the dimer and include all inter- and intralayer
two-body terms \cite{zinner2010}. However, here we are only interested
in the long-wavelength limit (momentum zero) in which the interlayer
term vanishes \cite{wang2008}. This gives
\begin{equation}
V_\textrm{eff}(U,\theta)=\frac{\hbar^2}{M}
\frac{4U}{3\sqrt{2\pi}}\left(\frac{d}{w}\right) 4\pi P_2(\sin\theta),
\end{equation}
where $P_2(x)=(3x^2-1)/2$. For the layer width, we take $w/d=0.2$ in
the following. Notice that $V_\textrm{eff}$ is attractive for
$\theta<\theta_c$, vanishes at $\theta_c$, and repulsive for
$\theta>\theta_c$. The attraction for $\theta<\theta_c$ results in a
negative compressibility in a single layer \cite{bruun2008}. We stress
again that $\theta_c$ is much smaller than the angle at which the
intralayer repulsion vanishes in a one-dimensional system. In this
sense $\theta_c$ is a special angle for the intralayer repulsion,
whereas it has no dramatic effect on the binding energies which vary
smoothly around $\theta=\theta_c$. Combining the formula above, the
final expression for $\mu$ becomes
\begin{equation}
\frac{Md^2}{\hbar^2}\mu=\frac{(k_Fd)^2}{2}\left(\frac{4U}{3\sqrt{2\pi}}
\left(\frac{d}{w}\right)P_2(\sin\theta)+1\right)-\frac{Md^2}{2\hbar^2}E_B,
\end{equation}
where we use the Fermi momentum $k_{F}^{2}=4\pi n$ for fermions in a single layer in place of $n$.

\begin{figure}
\centerline{\epsfig{file=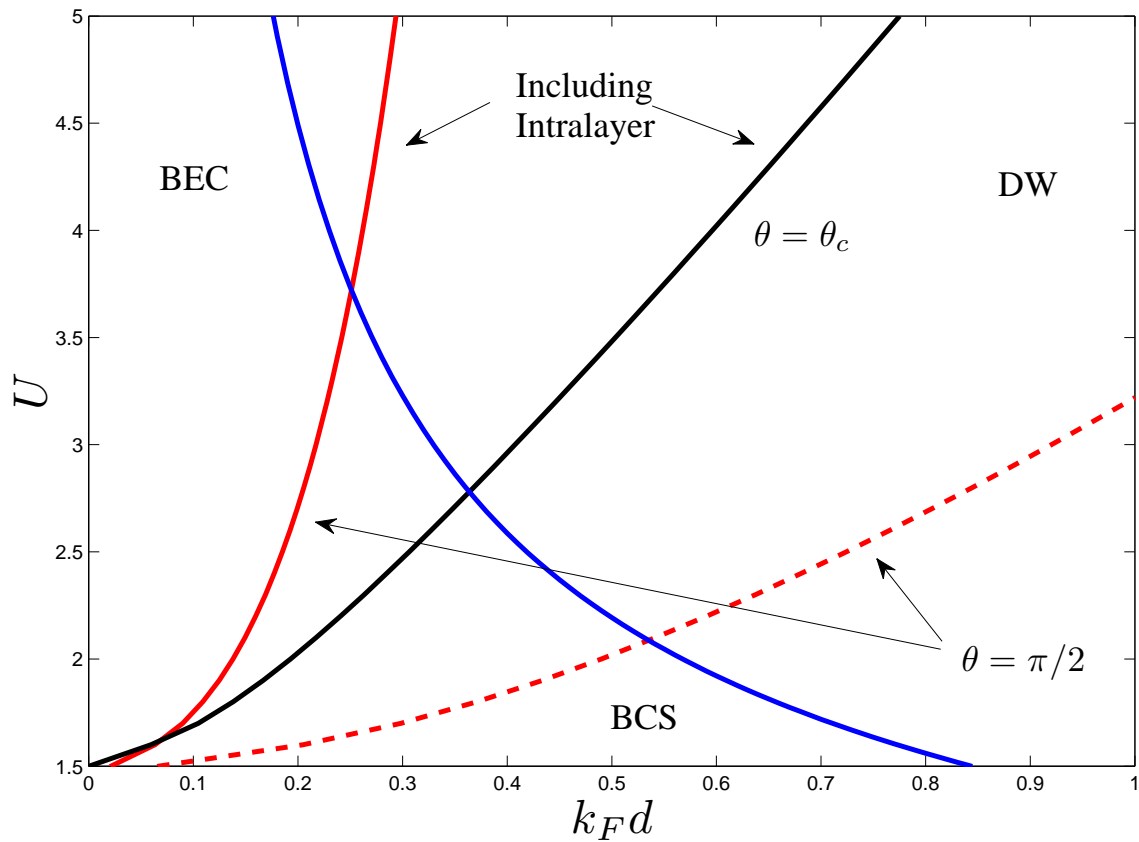,scale=0.75}}
\caption{Lines of vanishing chemical potential for
  $\sin^2\theta_c=1/3$ (solid black) and $\theta=\pi/2$ with (solid
  red) and without (dashed red) intralayer repulsion. Above the solid
  lines we expect condensation of dimers to occur (BEC), whereas in the
  lower right part a many-body paired BCS-like state (BCS) should be the
  ground-state of the system. Also shown is the line above which 
  a density wave (DW) for $\theta=\theta_c$ is expected in the bilayer 
  system (solid blue).}
\label{fig-strong}
\end{figure}

The lines of $\mu=0$ for selected angles are shown in
figure \ref{fig-strong} in the $(U,k_Fd)$ plane for $1.5<U<5$. 
For $U>2$ the
dimers have significant binding energy and can we treat them as localized bosonic
objects. For $\theta=\pi/2$ we present results both with and without
the intralayer term which is clearly seen to shrink the region of
potential dimer condensation. For $\theta=\theta_c$, the intralayer
term vanishes and we find a larger region of $\mu<0$. For
$\theta<\theta_c$ the region would in principle become even larger,
however, the intralayer term is then attractive and can lead to
instability and collapse \cite{bruun2008}. We therefore expect the
line for $\theta=\theta_c$ to provide a boundary for how large the BEC
region can become when tuning the angle within our approximations.
In figure \ref{fig-strong} we also show the line above which the 
density-wave instability appears at angle $\theta=\theta_c$ 
in the random-phase approximation 
as discussed in Ref.~\cite{zinner2011} (above full blue line, denote by DW).
To study the crossover outside the density-wave region, we see that 
low densities are indeed needed. 

We now consider the important question of the finite temperature behavior
of the system. In
the large $U$ limit, the BKT transition temperature is maximal at
$k_BT_{BKT}=E_F/8$ \cite{sarma2009,pikovski2010,zinner2010}, or
\begin{equation}
T_{BKT}=765\,\frac{n}{10^8\,\textrm{cm}^{-2}}\frac{\textrm{amu}}{M}\,\textrm{nK}.
\end{equation}
If we consider LiCs molecules (which can have a dipole moment of up to
5.5 Debye) with $d=0.5\,\mu$m, then the $\mu=0$ phase for
$\theta=\theta_c$ at $U=10$ is at $n=8.1\cdot
10^{7}\,\textrm{cm}^{2}$, and thus $T_{BKT}\sim 4.9$ nK. While still
very low, this is a significant increase over the sub-nano Kelvin
temperatures for $\theta=\pi/2$.

The $p$-wave superfluid for single
layers is estimated to appear for angles $\sin\theta<2/3$ \cite{bruun2008}.
However, density-waves occur for $\sin^2\theta>1/3$. 
This implies that there is a rather large regime ($2/3<\sin\theta\leq 1$)
where the single-layer $p$-wave superfluid should be absent but where
BCS-BEC crossover induced by the interlayer interaction is possible. 
We also note that there is an intermediate region where both $p$-wave
superfluid and density-wave have been proposed as the ground state and
one could imagine something like a supersolid phase as recently discussed 
in a two-dimensional optical lattice with fermionic polar molecules and
perpendicular polarization \cite{he2011}. In the strong-coupling limit
we expect that the allowed few-body states will play an important role
in determining the system properties, and calculation of the 
binding energies of states with three or more particles for arbitrary 
directions of the polarization are being pursued.

We expect that the partial-wave analysis presented in the previous 
section can help
indicate what symmetries are possible and relevant for the
corresponding many-body problem in the large $U$ limit. The problem is
of course still that the intralayer term is attractive in the
long-wavelength limit for $\theta<\theta_c$, and we thus expect that
the most stable system require $\theta>\theta_c$ where the
decomposition of the wave function is entirely dominated by the $m=0$
term. However, close to $\theta_c$ we still have a substantial $m=1$
contribution (immediately to the right of the vertical line in
figure \ref{fig-part}).  We therefore expect a region of interest in
which an exotic many-body state with non-trivial symmetry like a
$p$-wave dominated or mixed symmetry superfluid would emerge in the
bilayer. These indications are consistent with a partial-wave
decomposition of the potential itself \cite{mathy2010}. Combining this
information strongly suggests that there is a very interesting
crossover from weak- to strong-coupling in the corresponding many-body
system as recently discussed for the $\theta=\pi/2$ case
\cite{pikovski2010,zinner2010}. Similar considerations hold for the
$U<0$ case.

\section{Summary and Outlook}

We have studied a bilayer system of dipolar molecules for arbitrary orientation
of the dipoles with respect to the planes. The two-body bound state structure
was calculated, including energies and partial-wave decomposition of the 
wave function as function of dipolar strength and polarization angle.
We proved that there is always a bound two-body state in the system, 
irrespective of strength and polarization angle of the molecules, and
also verified this fact numerically. We 
argued that this follows from the fact that for small strength, the wave function
must reside outside the region where the potential is non-zero.
The results apply irrespective of the 
sign of the interaction strength. Negative strengths invert the dependence of energy
on the dipole angle such that perpendicular polarization angle has the smallest binding energy.
The structure of the wave function is dominated by the monopole component which 
decreases with the strength of the interaction. Up to moderate strengths, the 
monopole component is always larger than 50 percent while the dipole component
accounts for most of the remaining probability.

The conclusion that zero net volume potentials always have at least
one bound state could perhaps be reached in other ways. First,
approaching this limit from small negative net volume with one bound
state strongly indicates that the bound state remains. Second, a
perturbation argument is tempting, e.g. let us assume that the
cylindrical monopole potential always has a bound state (as shown
in \cite{simon1976}), and then treat
dipole and quadrupole terms as perturbations. We thus extend
from a Hilbert space entirely of $s$-waves to include also $p$ and
$d$-waves. To second order in perturbation theory this always gives
a negative contribution and hence more binding than for the monopole
potential alone.  However, closer scrutiny of such argements and their
practical implementation reveal that in the limit of weak potentials 
the perturbations are always of the same order as the initial
potential. This type of perturbative argumentation always fails. 
The deeper lying
reason is that the energy is a non-analytical function of the strength for
zero net volume potentials.  This is seen in the expression for the energy
where zero'th and first order in the strength contribute to the
non-analytical structure at the continuum threshold.  
Only terms higher than second 
order in the strength are perturbations around this strong singularity
which is a characteristic feature of zero volume potentials in two
spatial dimensions.

Implications for the many-body physics of a bilayer were discussed in the 
limit of strong-coupling where the two-body bound states are expected to 
be the important degrees of freedom. We conclude
that the region where (quasi)-condensation of two-molecule
dimers is likely to occur can be enhanced by tuning the angle of the dipoles. 
In particular, at the critical angle, $\sin^2\theta_c=1/3$,  
where the long-wavelength part of 
the intralayer interaction vanishes, we expect 
the conditions are most favorable for accessing this phase where dimers 
condense. This critical angle is different from the 'magic' angle
at which dipoles moving on a line become non-interacting which 
has been discussed in a number of previous works \cite{lahaye2009,santos2010,deu2010}.
We also estimated the potential for a density wave instability in the bilayer \cite{zinner2011}, 
and demonstrated that BCS-BEC crossover and (quasi)-condensation of 
dimers discussed here can occur also 
in an intermediate coupling and low density part of the
phase diagram which is below the density wave regime.
The possible roton instability in the bilayer system, as discussed for 
the perpendicular case in \cite{zinner2010}, is currently under active investigation.

The results presented in this paper indicate that bound complexes of more than 
two particles must exist in the bilayer. In the case of 
one-dimensional tubes this was studied in some special cases in
\cite{santos2010} and \cite{deu2010}, while a more complete
investigation of one-dimensional complexes as function 
of angles and dipole moment can be found in \cite{wunsch2011}. For 
the two-dimensional case, the method employed here can be
extended to complexes of more particles and we plan such investigations
in the near future. We also note that the external trapping 
potential that is present in each layer in experiments \cite{miranda2010} 
can be easily accomodated in the current approach by introducing 
one-body harmonic oscillator terms.

In conclusion, we find that bound states of dimers in a bilayer consisting
of one particle in each layer are generic for particles interacting
through the dipole-dipole force, irrespective of the dipole strength
or polarization angle of the dipoles with respect to the layers. In general, 
the wave function contains several partial-wave components and therefore
has interesting spatial structure. This suggests that few-body states with more
than two particles will also have rich structure and it also implies that
the many-body physics of the system is highly non-trivial. We sketched a 
phase diagram for the appearance of a dimer condensate as a function of 
polarization angle and showed that it is enhanced around the so-called 
magic angle. At this point the dimer contains a large admixture of 
higher partial waves and we expect the collective behavior of the 
system to reflect this fact. The many-body problem of a bilayer
with polar molecules of arbitrary polarization angle therefore deserve further
investigation.

\ack

We would like to thank J.~R. Armstrong, G.~M. Bruun, E. Demler, 
and C.~J.~M. Mathy for valuable discussions.

\appendix
\section*{Appendix}

\section{General derivation}
We assume reflection symmetry in the $x$-axis as for the potential in
equation (\ref{e20}) but we shall otherwise proceed with the general
derivation for any potential with this symmetry. The slightly more
general formulation without any symmetry can be found in \cite{vol11}.
We decompose the wave function into partial waves, $\cos(m\varphi)$,
i.e.
\begin{eqnarray}\label{eq-radial-functions}
&\Psi(s,\phi)= \frac{1}{\sqrt{s}} \sum_{m=0}^{\infty} 
 a_m \Phi_m(s)\cos(m\phi)
 &\\ & \frac{a_m \Phi_m(s)}{\sqrt{s}}=\frac{1}{(1+\delta_{m0})\pi}\int_{0}^{2\pi}
  d\phi \cos(m\phi) \Psi(s,\phi),&
\end{eqnarray}
where the corresponding contribution of the $\sin(m\phi)$ terms is
zero due to the assumed symmetry and the coefficients $a_m$ are real.
The normalization of the functions $\Phi_m(s)$ will be chosen later.
The rather artificial separation into coefficients and functions
allows a simple procedure to compute higher orders in the strength
$\lambda$.  The 2D radial wave functions, $\Phi_m(s)$, satisfy the
system of coupled radial equations
\begin{equation} \label{eq-w1} 
\Phi_m'' +\frac{1-4m^2}{4s^2}\Phi_m+\alpha^2\Phi_m
 = \lambda \sum_{l}\frac{a_l}{a_m}\Phi_l V_{ml}(s)\;,
 \end{equation}
where the matrix elements, $V_{ml}$, carrying all information about
the potential, are
\begin{equation} \label{eq-w2} 
V_{ml}=\frac{1}{(1+\delta_{m0})\pi} \int_0^{2\pi} \cos(m\varphi) \cos(l\varphi)
 \bar V(s,\varphi) \mathrm{d}\varphi \;.
\end{equation}
For cylindrical potentials $V_{ml} \propto \delta_{ml}$ and the
different $m$-values decouple.  The regular radial solution to
equation (\ref{eq-w1}) at the origin provides the usual boundary condition
for a centrifugal barrier potential, i.e. we choose the normalization
of $\Phi_m(\alpha,s)$ such that
$\lim_{s\to0}s^{-1/2-m}\Phi_m(\alpha,s)=1$.  We inserted here
explicitly the dependence on the energy parameter, $\alpha$, in
$\Phi_m$.  We assume here the potential, and consequently also the right
hand side of equation (\ref{eq-w1}), diverges slower than $1/s^2$ when
$s\rightarrow 0$.

We use the Green's function formalism which in 2D is described for a
cylindrical potential in \cite{new86}.  For anisotropic potentials we
have more generally that the $m$-components, $a_m\Phi_m$, are given by
\begin{eqnarray}
&a_m\Phi_m(\alpha,s)=a_m\Phi_{m0}(\alpha,s) \\&-
 \lambda \sum_{l}a_l\int_0^s\!g_{m}(\alpha,s,s')
 V_{ml}(s')\Phi_l(\alpha,s')\,\mathrm{d}s' \, . \label{eq-w4}
\end{eqnarray}
The last terms vanish for $s=0$ and the boundary condition at $s=0$
is obeyed through the first term:
\begin{equation} \label{eq-w5} 
\Phi_{m0}=\sqrt{s}J_{m}(\alpha s)(2/\alpha)^{m} m!\;, 
\end{equation}
where $J_m$ is the Bessel function.  The Green's function in the last
terms of equation (\ref{eq-w4}) is expressed as
\begin{eqnarray}  \label{eq-w9}
&&g_{m}(\alpha,s,s') =  \\ \nonumber
&&\frac{i\pi}{4}\sqrt{ss'} [H_{m}^{(1)}(\alpha s)H_{m}^{(2)}(\alpha s') 
 -H_{m}^{(1)}(\alpha s')H_{m}^{(2)}(\alpha s)] \;,
\end{eqnarray}
in terms of Hankel functions, $H_m^{(n)}$ \cite{abram64}.  At large
distance equation (\ref{eq-w4}), with the Green's function from equation (\ref{eq-w9}),
may be rewritten as
\begin{eqnarray}  \label{eq-w101}
&&a_m\Phi_{m}(\alpha,s)= \frac{1}{2}\sqrt{s}\left(\frac{2}{\alpha}\right)^m m! 
 \\ \nonumber && \times \left[H_{m}^{(1)}(\alpha s)\sum_l (c_{ml}^{*}+\delta_{ml})a_l +H_{m}^{(2)}(\alpha s)\sum_l (c_{ml}+\delta_{ml})a_l\right]\;,
\end{eqnarray}
where the matrix elements, $c_{ml}$, are:
\begin{eqnarray} \label{eq-w11}
&& c_{ml} \equiv  \\ \nonumber 
&& \lambda \left(\frac{\alpha}{2}\right)^{m}\frac{i\pi}{2m!} 
 \int_0^\infty\!\sqrt{s'}H_{m}^{(1)}(\alpha s') 
V_{ml}(s')\Phi_l(\alpha,s') \mathrm{d}s' \;.
\end{eqnarray} 
Equation (\ref{eq-w101}) contains incoming and outgoing waves and is applicable
for scattering problems.  For bound states, $\alpha$ has to be
imaginary corresponding to negative energy, and
${~}H_m^{(2)}(\alpha s)$ in equation (\ref{eq-w101}) diverges unless the
coefficient vanishes, i.e.
\begin{equation}\label{a3} 
\sum_{l=0}^{\infty} c_{ml}a_l=-a_m \;.
\end{equation}
This is an eigenvalue equation, which only has non-trivial solutions
for discrete values of $\alpha$, and hence for the binding energy.
However, the full radial wave functions, $\Phi_m$, enter in the
definitions of the matrix elements and the equations must be solved
selfconsistently.  We look for solutions in the limit of very weak
strength, $|\lambda|\ll 1$, which implies that $|\alpha|$ (i.e. the energy)
must also be very small.  The $m=0$ component will then dominate in
the solution to equation (\ref{a3}), because the centrifugal barrier
suppresses higher partial waves.  This suppression becomes more
pronounced with decreasing binding energy.

We now expand both coefficients, $a_m$, and functions, $\Phi_m$,
in powers of the strength, i.e. 
\begin{eqnarray} \label{a7}
a_m&=&\lambda a_m^{(1)}+\lambda^2 a_m^{(2)} + ...,  \\  \label{a8}
\Phi_m &=& \Phi_m^{(0)} + \lambda \Phi_m^{(1)} + \lambda^2 \Phi_m^{(2)}
+ ...,
\end{eqnarray}
where we leave the coefficient $a_0$ (without expansion) for
normalization of the total wave function.  In total we then find
\begin{equation} \label{a1}
\frac{a_m^{(1)}}{a_0}=-\frac{1}{2m}\int_0^\infty s^{1-m}V_{m0}(s)\mathrm{d} s \; ,
\end{equation}
\begin{eqnarray}
\label{a2a}
 \frac{a_m^{(2)}}{a_0}=-\frac{1}{2m}\int_0^\infty s^{\frac{1}{2}-m}V_{m0}(s)
 \Phi_0^{(1)}(0,s)\mathrm{d} s \\ - \frac{1}{2m}\sum_{i>0}\frac{a_i^{(1)}}{a_0} 
\int_0^\infty s^{\frac{1}{2}-m}V_{mi}(s)\Phi_i^{(0)}(0,s)\mathrm{d} s \; ,
\end{eqnarray}
\begin{eqnarray} \label{a4}
& \Phi_0^{(1)}(0,s) = - \sqrt{s} \int_0^{s} s'V_{00}(s') \ln (s'/s) 
 \rm{d} s' \;,\\  \label{a12} & \Phi_m^{(0)}(0,s) = s^{m}\sqrt{s} +
\frac{\int_0^{s} \frac{(s')^{2k} - (s)^{2k}}{(ss')^{m}} 
 s' V_{m0}(s') \rm{d} s'}{\int_0^{\infty} V_{m0}(s') (s')^{1-m} \rm{d} s'} \;,\;\;\;
\end{eqnarray}
\begin{eqnarray} \label{a2}
&& \Phi_m^{(1)}(0,s)=\int_0^s  g_m(0,s,s') \bigg( \sqrt{s'} 
\frac{a_0 a_m^{(2)}}{\left[a_m^{(1)}\right]^2}V_{m0}(s')- \nonumber \\&&
\frac{a_0}{a_m^{(1)}}V_{m0}(s')\Phi_0^{(1)}(0,s') -\sum_{i>0} 
 \frac{a_i^{(1)}}{a_m^{(1)}}V_{mi}(s')\Phi_i^{(0)}(0,s')\bigg) \mathrm{d} s'\; ,
\end{eqnarray}
\begin{eqnarray} 
 &&\Phi_0^{(2)}(0,s)= - \int_0^s \mathrm{d} s' g_0(0,s,s') 
  \label{a5}  \nonumber \\  && \times  \bigg(V_{00}(s')  \Phi_0^{(1)}(0,s')  
  + \sum_{i>0}\frac{a_i^{(1)}}{a_0}V_{0i}(s') \Phi_i^{(0)}(0,s')\bigg) \;,
\end{eqnarray}
\begin{eqnarray} 
 g_0(0,s,s') &=& \sqrt{s s'}\ln\frac{s'}{s} \; ,\\
g_m(0,s,s') &=& \frac{1}{2m}\sqrt{s s'}\frac{s'^{2m}-s^{2m}}{(s s')^m}  \;.
\end{eqnarray}
The equations \ref{a4},\ref{a2}, and \ref{a5} provide the expansions
for the coefficients and the wave functions in eqs. (\ref{a7}) and
equation (\ref{a8}) when $s\ll 1/|\alpha|$.  The behaviour of the wave function at infinity is now
given by the non-diverging piece in equation (\ref{eq-w4}), i.e.
\begin{equation} \label{eq-w6}
\lim_{s\to\infty}\Phi_{m}(\alpha,s) \sim\sqrt{s}H_{m}^{(1)}(\alpha s)\;, 
\end{equation}
or in the particular case of weak binding we have  
\begin{equation} \label{eq-w60}
\lim_{s\to\infty}\Phi_{m}(\alpha,s) \sim\sqrt{s}H_{m}^{(1)}(\alpha s)\delta_{m0}\;. 
\end{equation}
This behaviour is a consequence of the attractive and repusive
centrifugal barriers for $m=0$ and $m>0$, respectively.

The energy can now be found to any order in powers of $\lambda$. To
second order the results can be written as
\begin{eqnarray} \label{a9}
  E  &=& - \frac{2\hbar^2}{ \mu d^2} \exp(-2\gamma)  \\  \nonumber
 &\times& \exp\left( -\frac{2(1+\lambda B_1+ \lambda^2B_2)}
 {-\lambda I + \lambda^2(A_0 + \lambda A_1+\lambda^2A_2)}\right) \; ,
\end{eqnarray}
where the leading order constant is given by
\begin{eqnarray} \label{a11}
 I =\int_0^\infty s V_{00} \mathrm{d}s =\int_0^\infty \int_0^{2\pi} s
V(s,\varphi)\mathrm{d}s\mathrm{d}\varphi \;.
\end{eqnarray}
The potential dependent constants, $(A_i,B_i)$, become increasingly
more complicated with powers of $\lambda$.
When the net volume of the potential, $\lambda I < 0 $, is negative
the weak binding limit for an arbitrary potential is given by the
Landau expression \cite{landau1977}, which also turns out to be valid
for anisotropic potentials with the appropriate definition of the
volume. In the case of $\lambda I<0$, it is not necessary to retain
$A_2$ in (\ref{a9}) to second order in $\lambda$, whereas for $I=0$
it must be retained when calculating the corrections to the 
leading term.

When $I=0$ corresponding to zero net volume of the potential the
leading order term is given by $A_0$, i.e. 
\begin{eqnarray} \label{e128}
&A_0 \equiv  - \int_0^{\infty} \sqrt{s} V_{00}(s) \Phi_0^{(1)}(0,s) \rm{d} s + 
\\ \nonumber
 & \sum_{m \neq 0}  \int_0^{\infty} \frac{s^{1-m}}{2m} V_{m0}(s)  \rm{d} s 
 \int_0^{\infty} \sqrt{s'} V_{0m}(s') \Phi_m^{(0)}(0,s') \rm{d} s' \;, 
\end{eqnarray}
where only first order from equation (\ref{a1}) has to be used in the
derivation.

This derivation is completely general for two-dimensional,
anisotropic and reflection symmetric interactions.  The symmetry
requirement is only a minor simplification, and omitted in the
derivation in \cite{vol11}.  The overall results are that there is
always a bound state for very weak potentials with negative or zero
net volume, and the weak binding threshold behavior of the energy is
given by equation (\ref{a9}) to second order in the potential strength.
The leading order term for zero net volume is given by equation (\ref{e128})
where only the first term contribute for cylindrical potentials since
then $V_{0m}\propto \delta_{0m}$.  For non-cylindrical potentials even
the leading order expression in equation (\ref{e128}) is rather complicated.

Higher orders than those related to $I$ and $A_0$ are found for small energy
by an iteration procedure through eqs.(\ref{eq-w4}) and
(\ref{eq-w11}).  The radial solutions are computed from
equation (\ref{eq-w4}) which in turn are used to determine the $C$-matrix
and the energy.  This procedure can be repeated to give higher order
corrections of both energy and wave function. Much care is necessary
to include consistently all terms up to a given order because the
resulting expressions contain many terms. The simplest is $B_1$ which
is found to be
\begin{eqnarray} \label{e132}
B_1 &\equiv&  - \int_0^{\infty} V_{00}(s) \ln(s) s \rm{d} s \\ \nonumber
 &+& \sum_{m \neq 0}  \int_0^{\infty} \frac{s^{\frac{1}{2}-m}}{2m} V_{mm}(s) 
 \Phi_m^{(0)}(0,s) \rm{d} s  \;. 
\end{eqnarray}
The remaining expressions, $A_1,A_2,B_2$ are more complicated and we
do not show them here. They can be found by expanding $c_{00}$ in
equation (\ref{eq-w11}) up to fourth order in $\lambda$,
as well as $c_{0i}$ up to third order in
$\lambda$ in eqs.(\ref{a2}) and (\ref{a12}), and $c_{ml}$ up to second
order for $m \neq 0$ and $l \neq 0$.  Finally, we compute the
determinant of the matrix $c_{ml}+\delta_{ml}$, and equate to zero,
i.e.
\begin{eqnarray} \label{e133}
 && 1+ \lambda B_1 + \lambda^2 B_2 \\ \nonumber 
 && + \lambda^2 (A_0 + \lambda A_1 +
\lambda^2 A_2)  \ln(\frac{\alpha}{2} \exp(\gamma))  = 0  \;, 
\end{eqnarray}
which then directly leads to equation (\ref{a9}) with the identifiable
constants $A_1,A_2,B_2$.  If only $A_1$ is needed, lower orders are
required at each step but then $A_2,B_2$ are not obtained correctly.

\section{Matrix Elements}
The different matrix elements with shifted correlated $n$-dimensional
Gaussians can be calculated with the help of the following integrals,
\begin{equation}
\int d^nx e^{-x^TBx+v^Tx} =
\frac{\left(\sqrt{\pi}\right)^n}{\sqrt{\det{B}}}
\exp(\frac{1}{4}v^TB^{-1}v)\equiv \mathcal{M}\;,
\end{equation}
\begin{equation}
\int d^n x e^{- x^T B x + v^T x}  \left( a^T x \right)
= (a^T u) \mathcal{M}\;,
\end{equation}
\begin{equation}
\int d^n x e^{- x^T B x + v^T x}  \left( x^T D x \right)
= \left( u^T D u + \frac{1}{2} {\rm tr} (D B^{- 1}) \right) \mathcal{M}\;,
\end{equation}
\begin{eqnarray}
& \int d^n x e^{- x^T A' x + s'^T x} \left( -
\frac{\partial}{\partial x} \Lambda \frac{\partial}{\partial x} \right)
e^{- x^T A x + s^T x}  = \nonumber \\
& \left( 2 {\rm tr} (A' \Lambda A B^{- 1}) + 4 u^T A' \Lambda A u - 2
u^T
(A' \Lambda s + A \Lambda s') + s'^T \Lambda s \right)
\mathcal{M}\;,
\end{eqnarray}
where
\begin{equation}
B = A + A' ,\;
v = s + s' ,\;
u = \frac{1}{2} B^{- 1} v \;.
\end{equation}
For dipole-dipole interaction equation (\ref{e20}) the matrix elements of
with the long-range Gaussians equation (\ref{eq-long-range-gaussian}) are given as
	\begin{equation}
\int_0^\infty e^{-\frac{r^2}{b^2}}  V(r,\varphi) r
\mathrm{d}r\mathrm{d}\varphi=\frac{D^2}{d}
\pi\frac{3\sin^2(\theta)-1}{2}\left[
U(2,\frac{1}{2},\frac{d^2}{b^2})-2U(1,-\frac{1}{2},\frac{d^2}{b^2})
\right]
\;,
	\end{equation}
where
	\begin{equation}
U(a,b,z)= \frac{1}{\Gamma(a)}\int_0^\infty
e^{-zt}t^{a-1}(1+t)^{b-a-1}\mathrm{d}t\; ,
	\end{equation}
is the Tricomi confluent hypergeometric function. Note that only the monopole part, $V_0(r)$, gives
a non-zero contribution to the matrix element.

\end{document}